\documentclass[preprint]{aastex}%

\usepackage{amsmath}%
\usepackage{amsfonts}%
\usepackage{amssymb}%
\usepackage{graphicx}
\usepackage{multirow}
\usepackage{subfigure}
\usepackage[english]{babel}
\usepackage[utf8]{inputenc}

\begin{document}

\title{Transit model of planets with moon and ring systems}

\author{Luis Ricardo Moretto Tusnski}
\affil{Astrophysics Division, Instituto Nacional de Pesquisas Espaciais \\ Av. dos Astronautas, 1758, Sao Jose dos Campos, SP, Brasil.}
\email{lrtusnski@das.inpe.br}

\and

\author{Adriana Valio}
\affil{Center for Radio Astronomy and Astrophysics Mackenzie, Universidade Presbiteriana Mackenzie \\ Rua da Consolacao, 896, Sao Paulo, SP, Brasil.}
\email{avalio@craam.mackenzie.br}

\begin{abstract}

Since the discovery of the first exoplanet, those most adequate for life to begin and evolve have been sought. Due to observational bias, however, most of the discovered planets so far are gas giants, precluding their habitability. If these hot Jupiters are located in the habitable zone of their host stars, and if rocky moons orbit them, these moons may be habitable. In this work we present a model for planetary transit simulation considering the presence of moons and planetary rings around the planet. The moon's orbit is considered to be circular and coplanar with the planetary orbit. The other physical and orbital parameters of the star, planet, moon and rings can be adjusted in each simulation. It is possible to simulate as many successive transits as desired. Since the presence of spots on the surface of the star may produce a signal similar to that of the presence of a moon, our model also allows for the inclusion of starspots.
The result of the simulation is the light curve with the planetary transit. White noise may also be added to the light curves, to produce curves similar to those obtained by the CoRoT and Kepler space telescopes. The goal is to determine the criteria for detectability of moons and/or ring systems using photometry. The results show that it is possible to detect
moons with radii as little as 1.3 R$_\earth$ with CoRoT and 0.3 R$_\earth$ with Kepler. 

\end{abstract}


\section{Introduction}

Since the detection of the first extra-solar planet orbiting a solar-like star in 1995, 51 Peg b \citep{Mayor1995}, several new discoveries have been made. The most used planetary detection methods are radial velocity and transit. Together, both techniques led to the discovery of the vast majority of the over 570 planets observed so far (according to the Extra-solar Planets Encyclopaedia, http://exoplanet.eu).  When the same planet is detected by both methods, the complete set of planetary and orbital parameters is determined. Two space missions designed to detect planetary transits are operational right now: CoRoT, from the CNES with a consortium of many countries, and Kepler, from NASA.

One of the motivations for the search of extra-solar planets is the quest for places suitable for the emergence of life. For this to be possible, the planet needs to be located in a place known as the habitable zone \citep{Huang1959, Kasting1993}. This is the region around the host star, where the radiation received by the planet is such that allows for water to exist in its liquid state. Furthermore, the planet needs to be rocky. Due to the observations bias, the majority of the planets found up to date are gaseous giants, what excludes them from being habitable even if they are in the habitable stellar zone.

Through observation of the Solar System, we can infer that the existence of moons and planetary rings around gas exoplanets may be common. Also, we can deduce that moons are rocky by nature. Planetary rings may exist in very different configurations, from the rarefied rings of Uranus to the giant rings of Saturn. The detection of both moons and/or rings may uncover important clues about the planet formation and migration process. Moreover, if a gas planet located in the stellar habitable zone has moons, these moons may also be habitable, making them a good place for life to evolve.

\cite{Sartoretti1999} suggest that exomoons could be detected using photometry. The presence of moons around an exoplanet would cause distortions in the transit light curve, in the form of steps or asymmetries. The detection of these distortions would provide the moon's orbital parameters. Using this technique and photometric data from the Hubble Space Telescope, \cite{Brown2001} searched for moons and rings around the exoplanet HD 209458b. However, nothing was found. \cite{Kipping2011} proposes an algorithm capable of generating transit light curves of a planet with a moon analytically, and shows how this can be used on the search for moons using photometry. On the other hand, \citet{Han2002} studied the detection possibility using gravitational micro-lensing events. \citet{Williams2004} showed that it is possible to detect moons in orbit of gaseous giant planets in the habitable zone using spectroscopy in the 1.5 - 5 $\mu$m band. \citet{Kipping2009, Kipping2009b} suggests that moon detection is possible from the effects on the planet motion, that may be detectable through precise measurements of the timing and duration of the transits. These are known as \emph{transit time variation} (TTV) and \emph{transit duration variation} (TDV).

\cite{Barnes2004} suggest that rings around a Saturn-like exoplanet could be detected with photometric precision of (1 - 3) $\times$ 10$^{-4}$ and time resolution of 15 minutes, as long as the ring is not viewed close to edge-on. This resolution is within the accuracy that the Kepler Space Telescope can achieve for Sun-like or brighter stars. \cite{Ohta2009} study ring detection through photometry and spectroscopy, and conclude that rings with significantly great inclination can be detected by today radial velocity precision of 1 m/s. \cite{Schlichting2011} study the nature of rings that can exist around exoplanets, and conclude that rings around close-in planets can show non-trivial Laplacian planes. This is caused by the increased effects of the planet's quadrupole moment. These so called warped rings furnish important clues about the planet interior and spin period.

Despite all these proposed methods, no exo-moon or planetary rings have been observed so far. 

Here, we propose a model of planetary transit simulation that may be used in the search for exo-moons through their signature in the light curve of their host planets. Beside the moons, the model may also simulate the transit of a planet with a ring around it. The model we propose here is capable of generating transit light curves numerically, unlike \cite{Kipping2011}. Albeit the larger processing time needed to fit light curves using this method, it has the advantage of being easily adaptable to include new features, as additional moons, changes in the ring plane or starspots. The goal is to apply this model to the observations of the CoRoT and Kepler space telescopes, looking for the photometric signals that may indicate the presence of moons or rings around exoplanets. 

Our model does not consider the detection of the warped rings shown by \cite{Schlichting2011}. Such rings are predicted around close-in exoplanets, with orbital period of a few days. Thus, our model is limited to the detection of rings located in the planet's equatorial plane, with larger semi-major axis. As the occurance of such warped rings depends on the quadrupole moment J$_{2}$ and on the density of the planet, it is difficult to stablish a general criterion to the validity of our model. For example, for the case of an exoplanet with density 3 g cm$^{-3}$ and J$_{2}$ = 10$^{-2}$, warped rings occur for semi-major axis below 0.05 AU (see Figure 6 on \cite{Schlichting2011}). Planets with larger semi-major axis will have rings located in the planet equatorial plane and could be detected by our model.

The next section describes the model used in this work, whereas Section 3 presents the application of this model to a few study cases. Section 4 discusses the  detectability threshold of moon and ring systems. Finally, the last section lists our main conclusions.

\section{The model}

The model used here is based on that of \citet{Silva2003}, where the star is considered a disc with limb darkening and the planet a completely dark disc. Both stellar and planetary parameters can be fit by the model. To minimize the number of free parameters, we consider a simple model where the moon's orbit is coplanar with the planetary one, and is also circular. Looking at the Solar System's moons, we see that these assumptions are a little limiting. However, as this is our first study on the detection of moons, we choose to consider the simplest model. In a future work we can consider a more realistic model, including orbital inclinations and eccentricity for the moon. Regarding to the planet eccentricity, following the strategy proposed by \cite{Kipping2008}, the model simulates eliptical orbits only for the cases were eccentricity was measured by other methods, such as radial velocities.

Stellar input parameters to the model are: rotation period, radius, and mass (the latter two in solar units). The rotation period is used in the calculation of the longitude position of starspots in successive transits. We also have two limb darkening parameters. The planetary input parameters are: orbital period, radius, mass, orbital inclination angle, eccentricity and position of the periastron. From the orbital period, the orbit's semi-major axis is calculated by Kepler's Third Law. Lastly, the moon input parameters are orbital period, radius, mass, and the angular position of the moon in its orbit at the starting time of the first transit. For the rings, the input parameters consists of inner and outer radii, two inclination angles, one with respect to the plane of the sky and the other with the planet's orbital plane. The transparency of the rings may also be fit.

The temporal resolution of the resulting light curve is chosen within the model, by adjusting the time interval for each simulated photometric data point. Here we used a temporal resolution of 32 s for future comparison with CoRoT data, and 1 min for Kepler data. 

Using stellar and planetary parameters, the planetary orbit is calculated. The line of transit is the projection of this orbit on the stellar surface. At each time interval, the program calculates the position of the planet and its moon. The total luminosity of the star--planet--moon system is then determined by summing the intensities of all the pixels in the image, where the pixels occupied by the planet and its moon have zero value. In the case of rings, the occupied pixels intensity is multiplied by the ring transparency factor. At the end of the transit, the light curve is normalized to one from the outside transit data points.

\begin{figure*}[!htb]
   \centerline{
     \subfigure[ ]{\includegraphics[width=5.5cm]{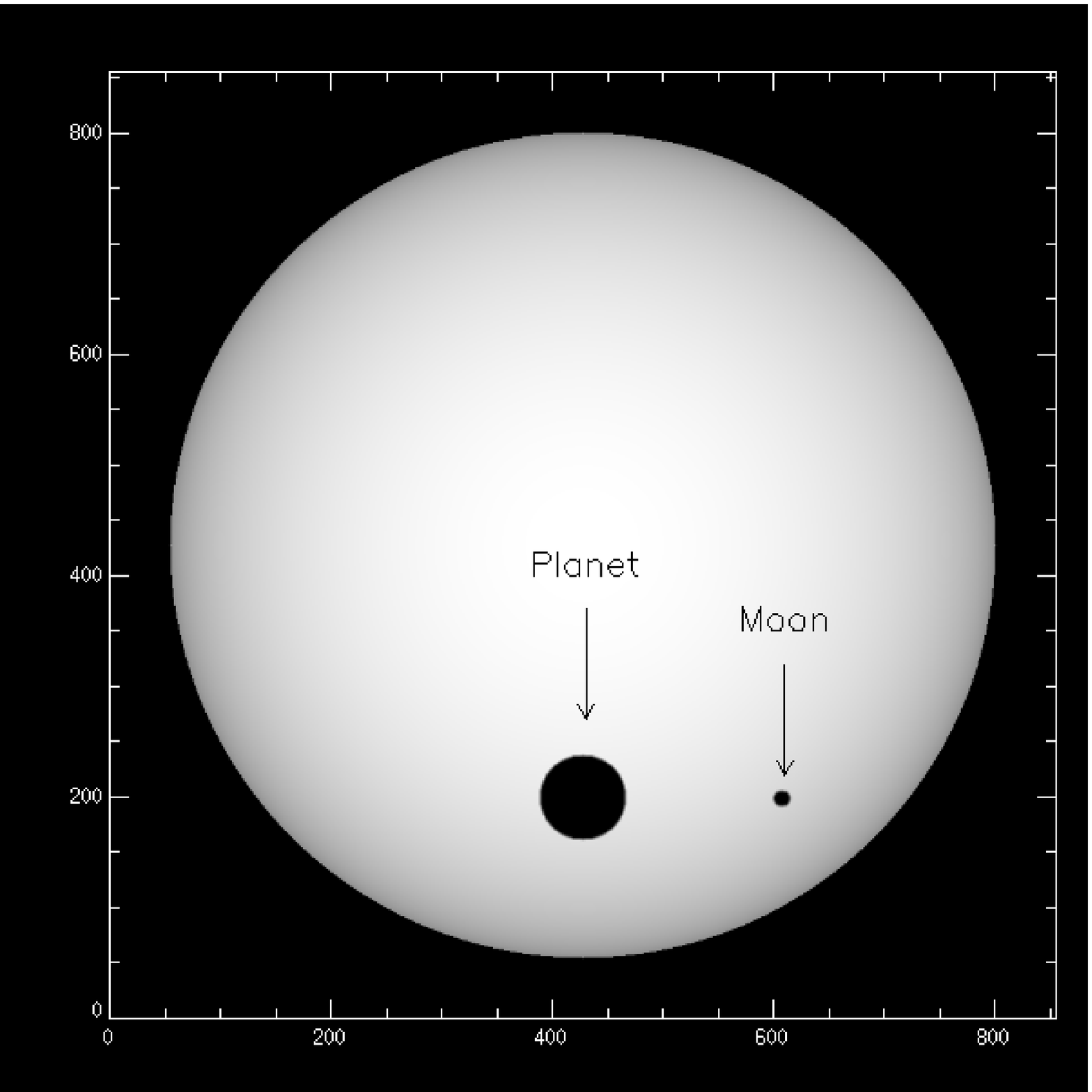}
     \label{fig:img_transitos_a}}
     \hfil
     \subfigure[ ]{\includegraphics[width=5.5cm]{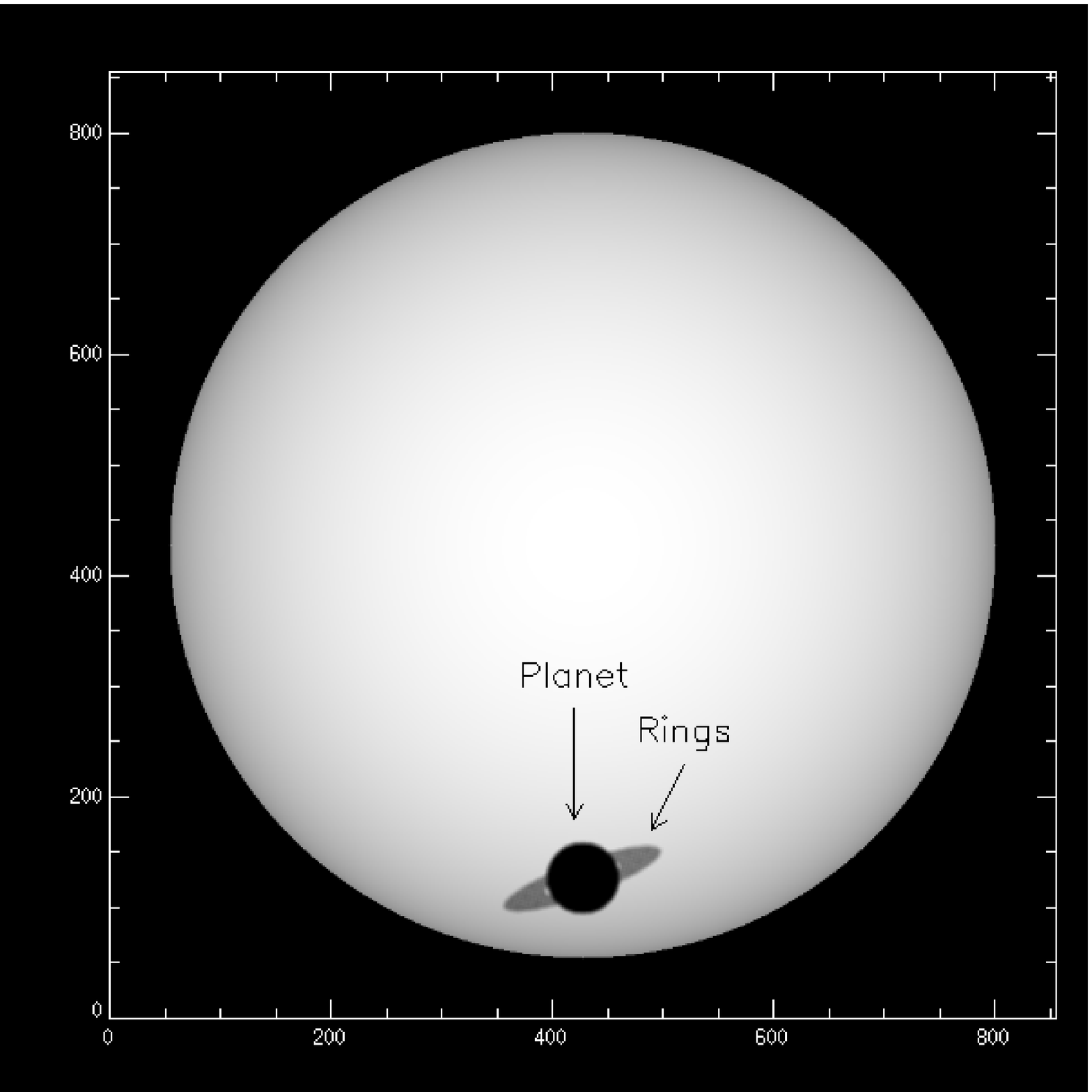}
     \label{fig:img_transitos_b}}
   }
   \caption{\textbf{(a)} A Jupiter-like planet and a 2 R$_{\oplus}$ moon transiting in front of the star. \textbf{(b)} A Saturn-like planet with rings transiting in front of the star. The rings have the same dimensions as Saturn's. The angle with the sky plane is 78$^\circ$, and the angle with orbital plane is 20$^\circ$. Both images were obtained from the model.}
   \label{fig:img_transitos}
\end{figure*}

Figure \ref{fig:img_transitos} shows two examples of transits from the model. Figure \ref{fig:img_transitos_a} shows the transit of a planet and a moon in front of the star, and Figure \ref{fig:img_transitos_b} shows the transit of a planet with rings.

It is also possible to add spots on the stellar surface, where each spot is modeled by three extra parameters: radius, intensity (with respect to stellar central intensity), and longitude position at the start of the first transit. The spot latitude is considered to be that of the projected transit line. From the stellar rotation period, the spot longitude position on consecutive transits is calculated. A further assumption is that the spot position does not change within a single transit (usually lasting a few hours), which is reasonable when the stellar rotation period is much larger than the transit duration.

\subsection{Planet-Moon dynamics}

In this work, no attempt has been made for a dynamical analysis of a planet with moons or rings. We used for this matter, regarding to planet-moon systems, the results of \cite{Domingos2006}, where the authors through numerical simulations obtain semi empirical expressions for the semi-major axis maximum and minimum that a satellite in orbit of a giant extrasolar planet may have as a function of the stellar and planetary parameters. \cite{Domingos2006} also estimates the maximum mass of a satellite such that its orbit is stable. An interesting result is that planets with very close in orbits to their host star can only hold moons with very small masses, but more distant planets may have stable moons with masses even larger than that of the Earth.

For example, \emph{Rho Crb} at a distance of 0.224 AU from its host star, with an orbital period of 39.8 d, may harbor a satellite of up to 1.5 M$_\oplus$. Further away planets, such as HD 92788, at approximately 1 AU from its star, allows for a satellite with a maximum mass of $1 \times 10^4$ M$_\oplus$. The transiting extrasolar planets with the longest periods observed so far are CoRoT-9b (95.27 d), HD 80606 (114.34 d) and Kepler-11 g (118.37 d). Thus, these planets may harbor a moon with mass similar or even larger than Earth, that may be detectable from the application of the present method to CoRoT or Kepler lightcurves.

\subsection{Planetary transit} \label{secao_transito_planeta}

\begin{figure*}[!htb]
   \centerline{
     \subfigure[ ]{\includegraphics[width=5.5cm]{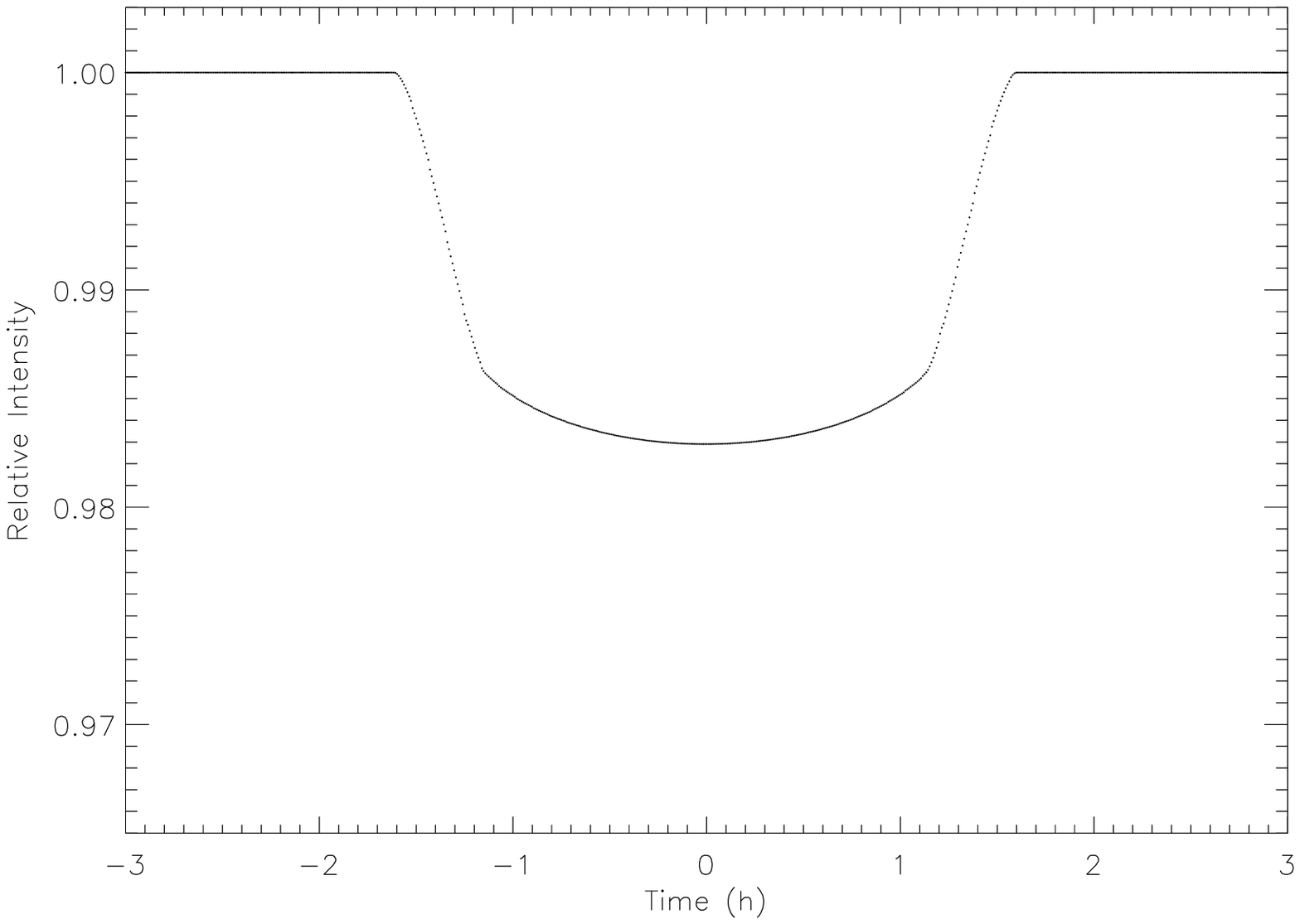}
     \label{fig:planetas_sem_lua_a}}
     \hfil
     \subfigure[ ]{\includegraphics[width=5.5cm]{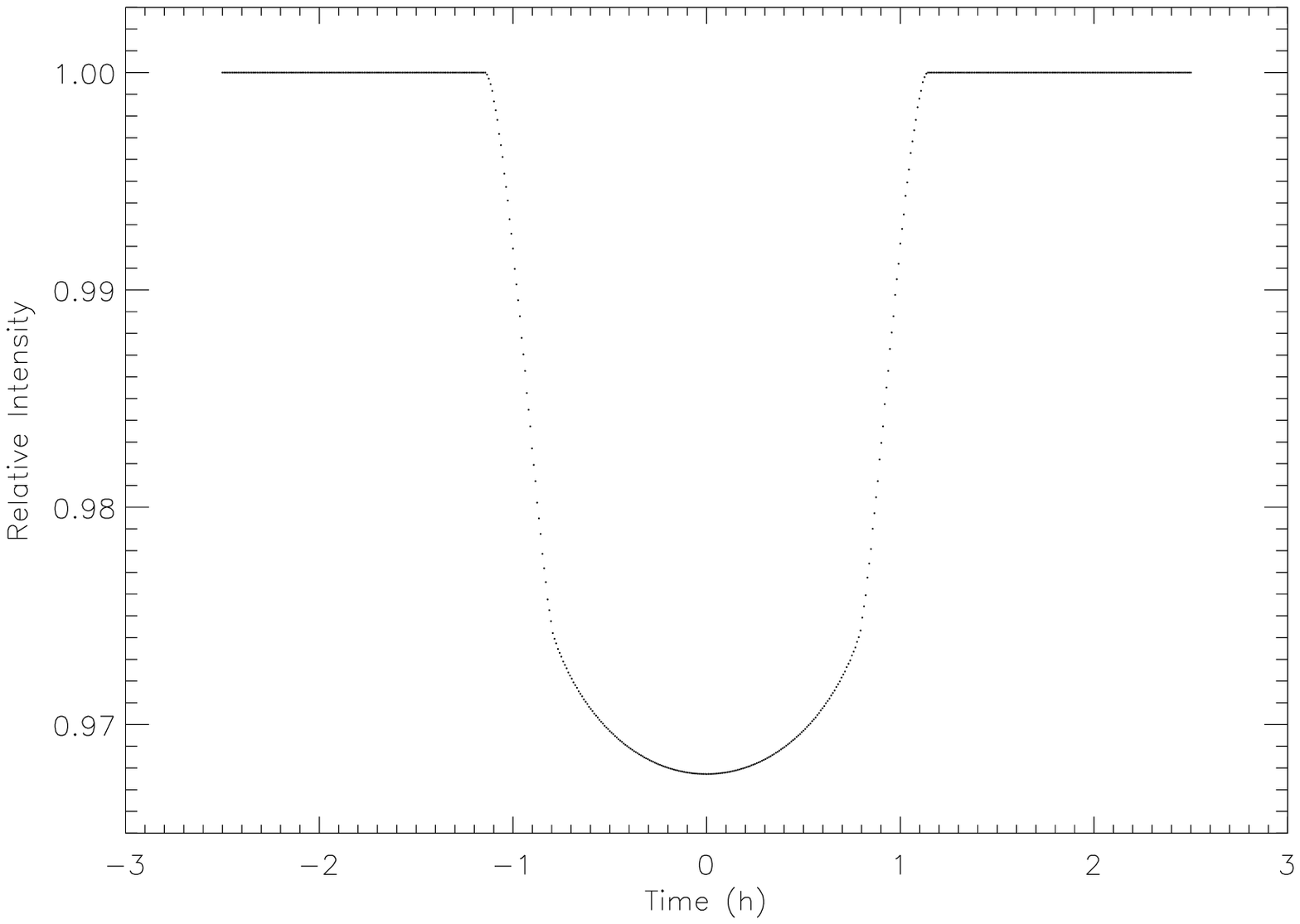}
     \label{fig:planetas_sem_lua_b}}
   }
   \caption{Transit model of two planets. \textbf{(a)} HD 209458b (R = 1.38 R$_{\mbox{J}}$, P = 3.524 days, i = 86.59$^\circ$), and \textbf{(b)} CoRoT-2b (R = 1.465 R$_{\mbox{J}}$, P = 1.743 days, i = 87.84$^\circ$).}
   \label{fig:planetas_sem_lua}
\end{figure*}

Figure \ref{fig:planetas_sem_lua} shows two examples of lightcurves obtained from our model. The figure on the left, \ref{fig:planetas_sem_lua_a}, is for the transit of a planet like HD 209458b (\cite{Southworth2010}), whereas the figure on the right, \ref{fig:planetas_sem_lua_b}, is for a planet similar to CoRoT-2b (\cite{Alonso2008}).

\subsection{Transit of a planet with a moon}

\begin{figure*}[!htb]
   \centerline{
     \subfigure[ ]{\includegraphics[width=5.5cm]{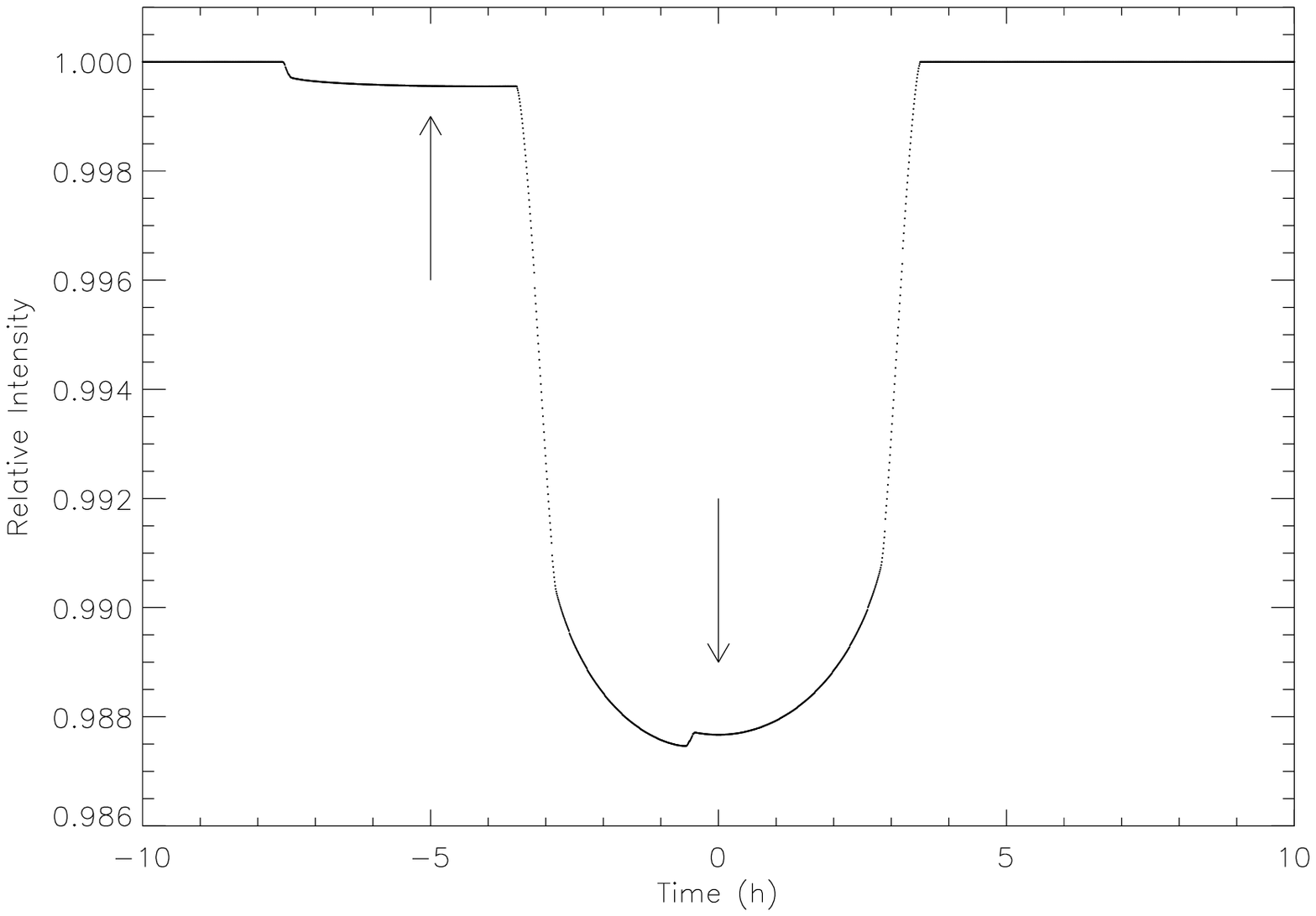}
     \label{fig:planetas_com_lua_a}}
     \hfil
     \subfigure[ ]{\includegraphics[width=5.5cm]{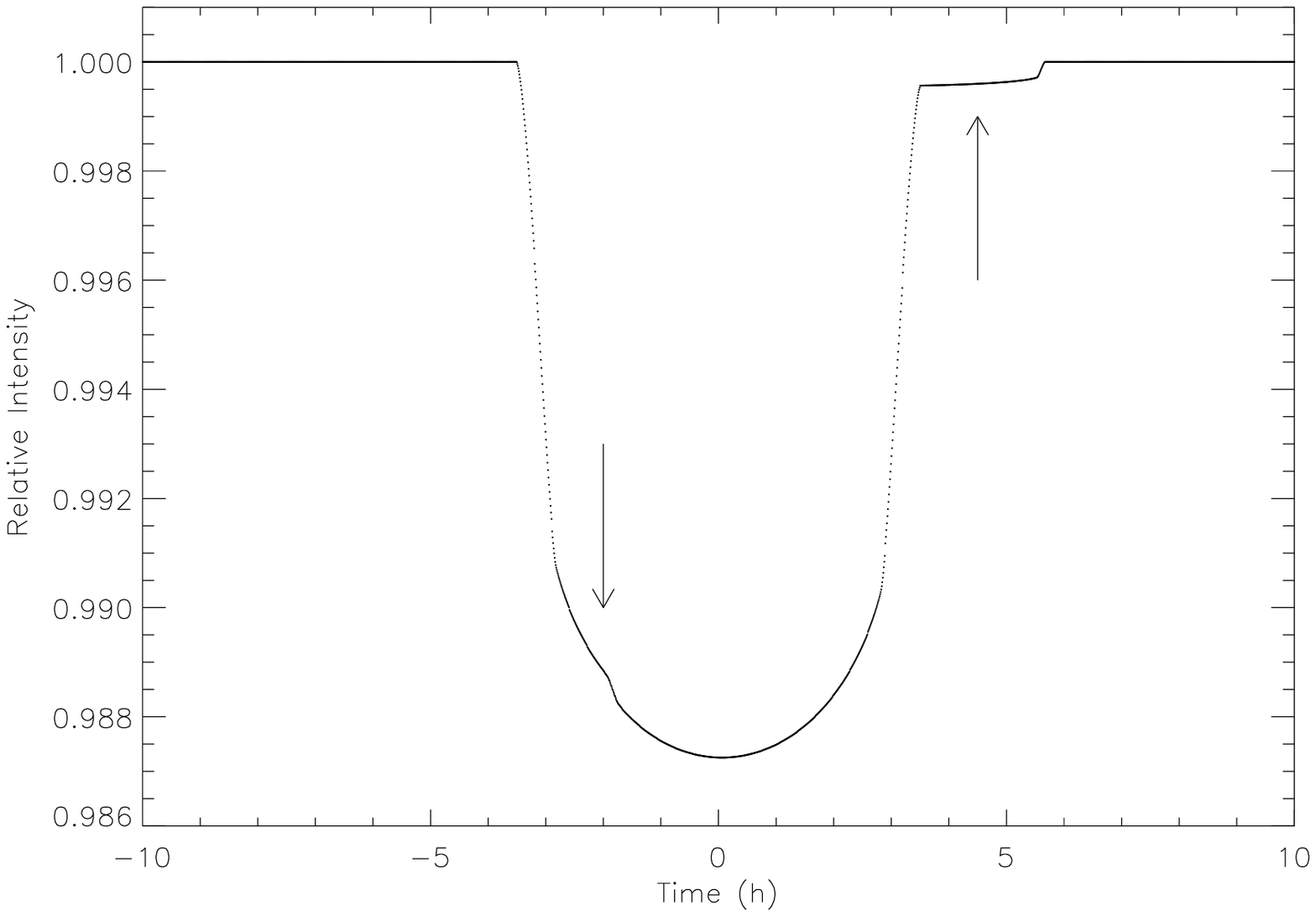}
     \label{fig:planetas_com_lua_b}}
   }
   \caption{Model light curve of a transiting hot Jupiter (R = 1 R$_{\mbox{J}}$, P = 45 days, i = 89.5$^\circ$) around a Sun-like star, and an orbiting moon with 2 R$_{\oplus}$. The moon period and position angle are, respectively: \textbf{(a)} $P_M = 5$ days and $\varphi = 0^\circ$; \textbf{(b)} $P_M = 2$ days and $\varphi = 60^\circ$.}
   \label{fig:planetas_com_lua}
\end{figure*}

Figure~\ref{fig:planetas_com_lua} shows two examples of lightcurves of a planet with an orbiting moon. Both panels are for a Jupiter size planet transiting a star identical to the Sun with an orbital period of 45 days. In the model, the added moon around this planet has 2 Earth radii. In \ref{fig:planetas_com_lua_a}, the moon orbital period is 5 days, while in Figure~\ref{fig:planetas_com_lua_b} it is 2 days. Also, the position angle of the moon with respect to the planet is $0^\circ$ in the first case, and  $60^\circ$ in the second one. 

When adding a moon around a planet in the model, two characteristics signatures may occur. The first one is a step, that may appear at the beginning or the at the end of a transit, while the second effect is a discontinuity, a jump, seen in the bottom of the transit. 
The step at the beginning (see Figure~\ref{fig:planetas_com_lua_a}) occurs when the moon ingress happens before that of the planet, while the step near the end of the transit (Figure~\ref{fig:planetas_com_lua_b}) is caused by the moon egress after the planet. Both effects are marked with an upward arrow.

Furthermore, when the moon orbital period is small, an asymmetry arises in the bottom of the transit, shown by the downward arrows in Figure~\ref{fig:planetas_com_lua}. This asymmetry is generated when the moon is eclipsed by the planet. It may also happen if at the start of the transit the moon is already behind (or in front) of the planet, but falls out of eclipse during the transit. The width and intensity of this signal depends on the moon's relative orbital period, radius, and also on the position angle $\varphi$ of the moon with respect to the planet at the initial transit time.

\subsection{Transit of a planet with rings}

\begin{figure*}[!htb]
   \centerline{
     \subfigure[ ]{\includegraphics[width=5.5cm]{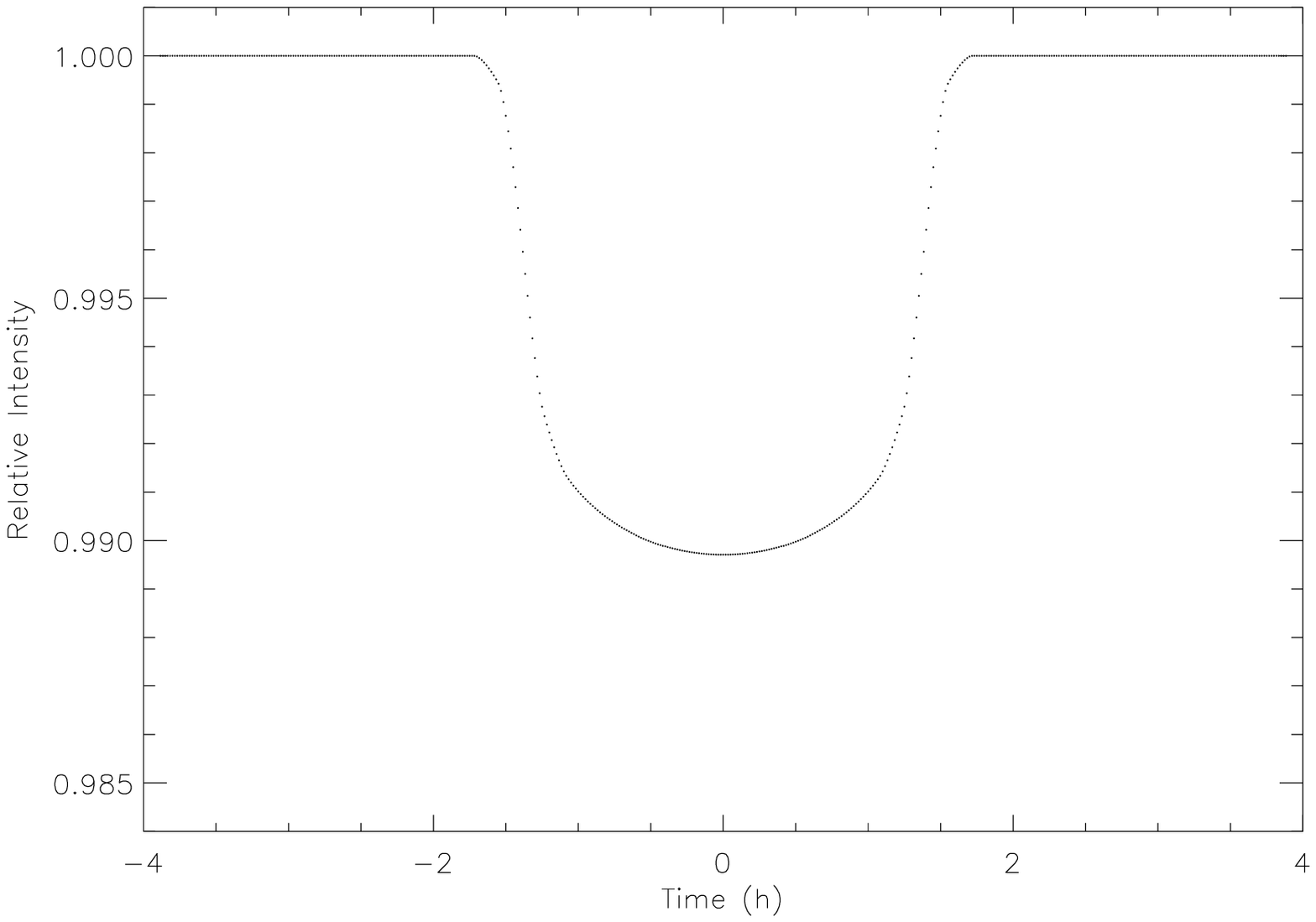}
     \label{fig:planetas_com_aneis_a}}
     \hfil
     \subfigure[ ]{\includegraphics[width=5.5cm]{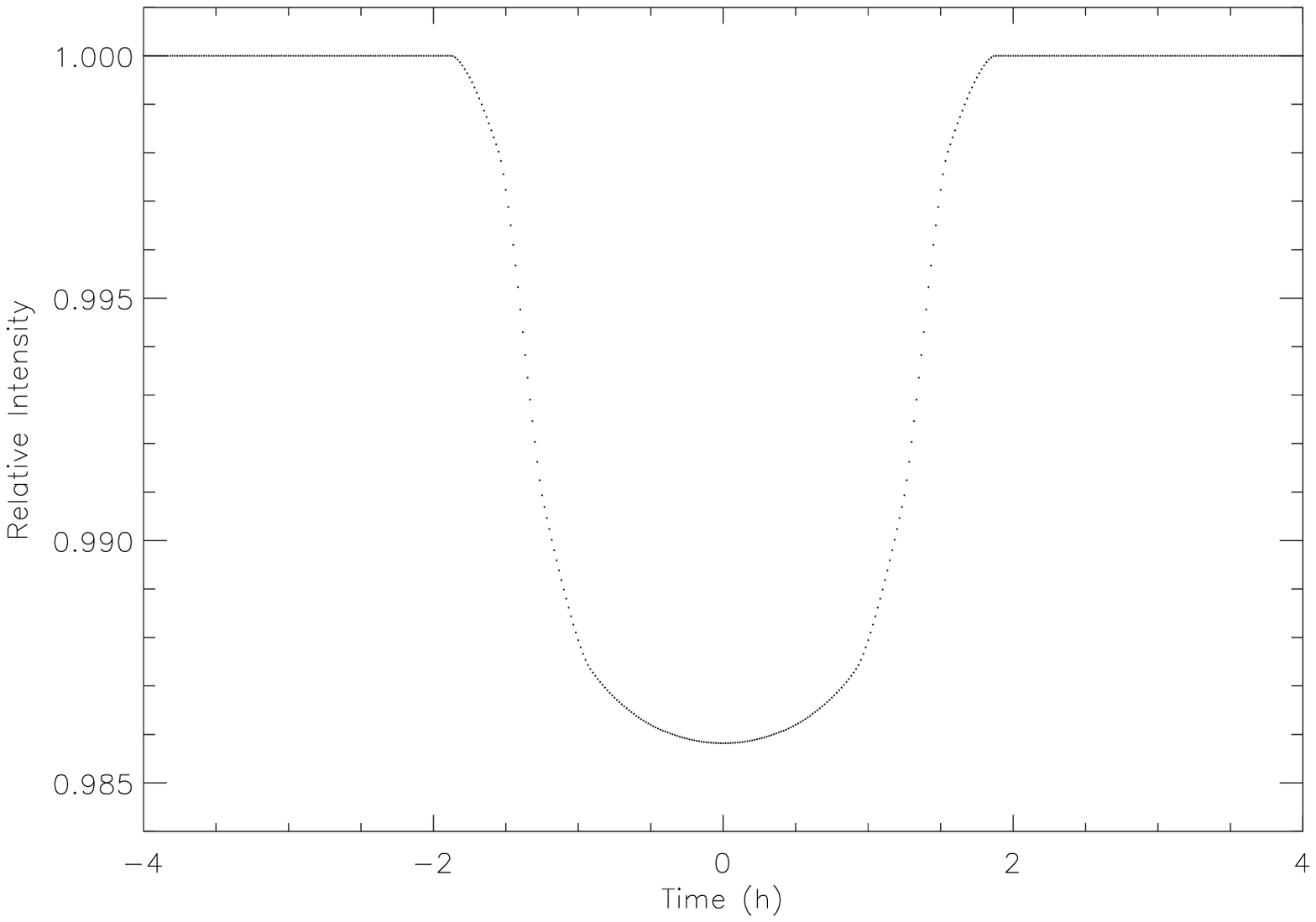}
     \label{fig:planetas_com_aneis_b}}
   }
   \caption{Transiting Saturn-like planet (R = 0.84 R$_{\mbox{J}}$, P = 5 days, i = 88$^\circ$) in front of a solar-like star.  Planetary rings with sizes \textbf{(a)} equal and \textbf{(b)} increased by 50\% with respect to those of Saturn.}
   \label{fig:planetas_com_aneis}
\end{figure*}

Figure~\ref{fig:planetas_com_aneis} presents the modeled light curve produced by a transiting Saturn-like planet around a star with solar parameters (left panel). To enhance the signal produced by the ring presence, their size were augmented by 50\% (right panel).

The main effect of the planetary ring system is to increase the covered area of the stellar surface. The result of which is to deepen the transit light curve. Moreover, the transit shape becomes rounder at the ingress and egress instants, than if no rings are present. Both these factors depend strongly on the ring transparency, being larger in the case of more opaque rings, as expected.

An interesting effect appears when the ring size is increased. In this case, the light curve becomes more triangular, similar to the light curves produced during grazing eclipses of binary star systems.

\subsection{Starspots}

If the star has one or more spots on its surface, similar to sunspots,  \citet{Silva2003} has shown that the photometric light curve during a planetary transit will display a small ``bump". This occurs because when the planet occults a darker region of the stellar surface, occupied by the spot, there is a relative increase in the total intensity. Moreover, the detection of the same starspots on different transits may be used to determine the rotation period of the star (\cite{Silva2008}).

\begin{figure*} [!htb]
   \centerline{
     \subfigure[ ]{\includegraphics[width=5.5cm]{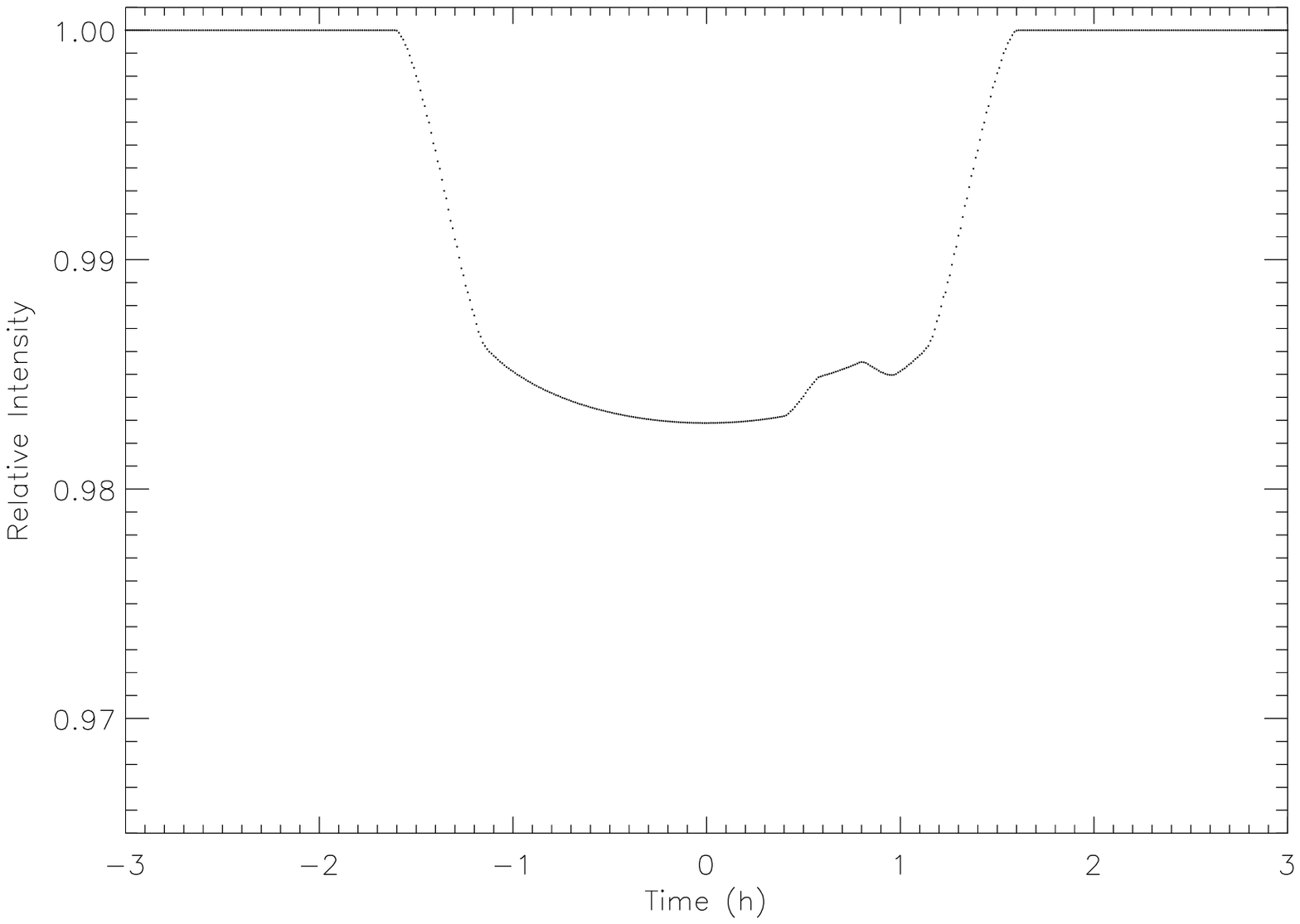}
     \label{fig:planetas_com_mancha_a}}
     \hfil
     \subfigure[ ]{\includegraphics[width=5.5cm]{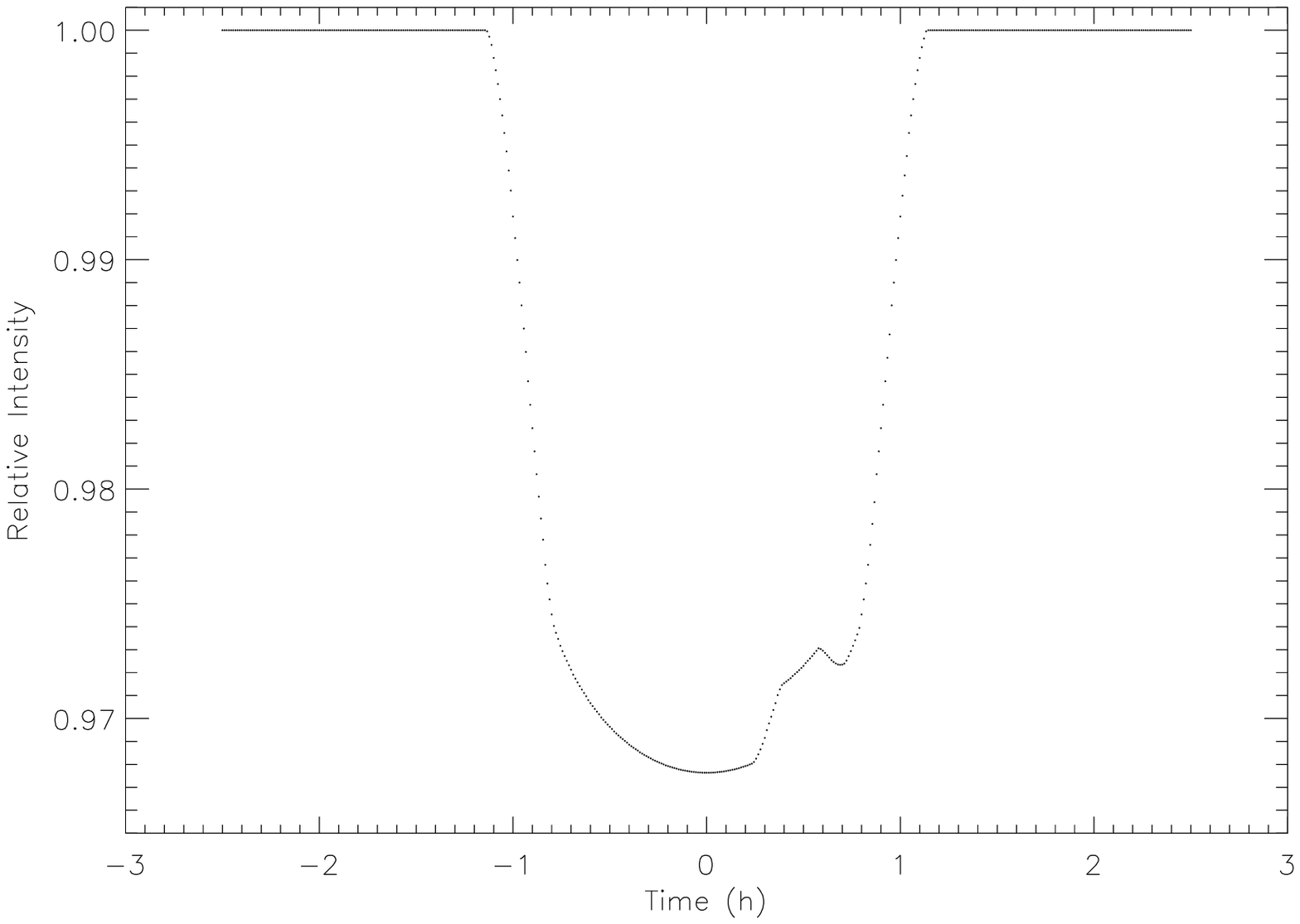}
     \label{fig:planetas_com_mancha_b}}
   }
   \caption{Planetary transit light curves for \textbf{(a)} HD 209458b and \textbf{(b)} CoRoT-2b, where the ``bump" in the bottom of the transit indicates the presence of a spot on the stellar surface. In each case, the spot has half the planet radius and intensity 0.6, and is located at 30$^\circ$ with respect to the star center. }
   \label{fig:planetas_com_mancha}
\end{figure*}

Figure~\ref{fig:planetas_com_mancha} shows the same light curves from Figure~\ref{fig:planetas_sem_lua}, except now a spot was added to the stellar surface. This spot, located at 30$^o$ longitude (measured from the center of the star), has a size of half each planet radius and a relative intensity of 0.6 times the stellar central intensity. 

In some cases, the photometric signal produced by the presence of moons may be mistaken by that caused by starspots. Distinguishing between the two cases is possible when successive transit are observed, since the spots signal will have a periodicity equal to the stellar rotation period, while those of the moons will have a different period, following its own orbital period.

\begin{figure*}[!htb]
   \centerline{
     \subfigure[ ]{\includegraphics[width=5.5cm]{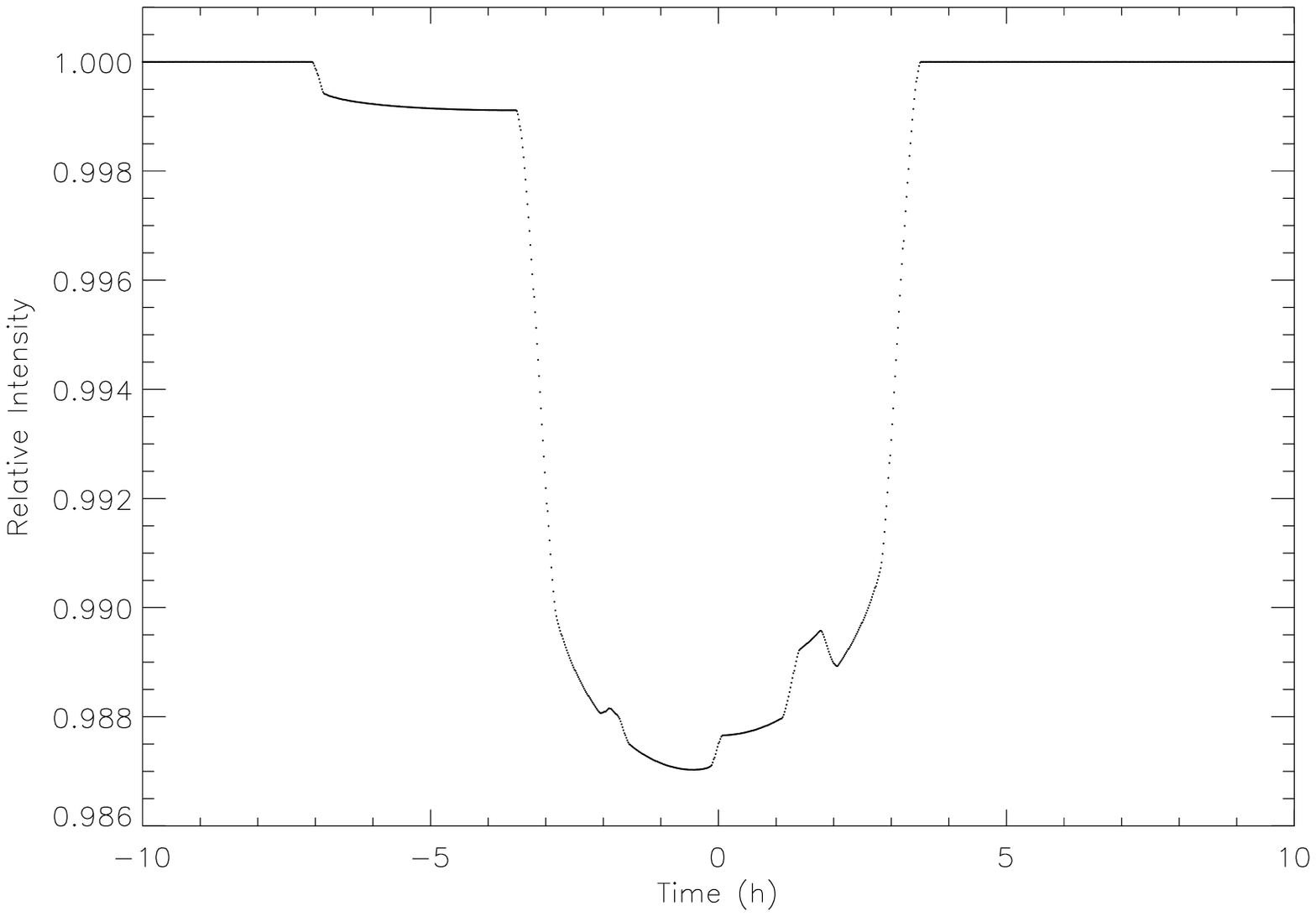}
     \label{fig:planetas_com_lua_e_manchas_a}}
     \hfil
     \subfigure[ ]{\includegraphics[width=5.5cm]{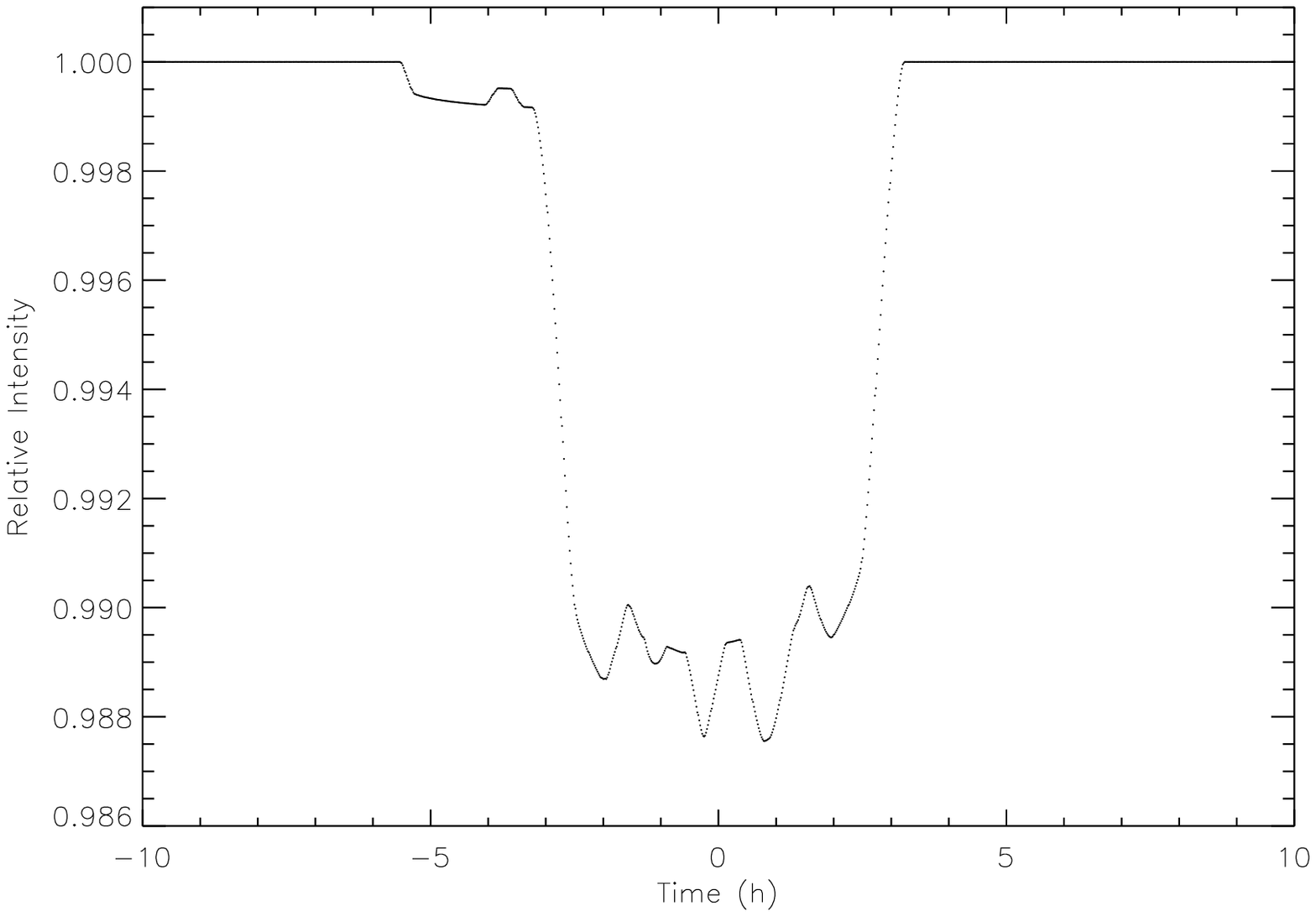}
     \label{fig:planetas_com_lua_e_manchas_b}}
   }
   \caption{Jupiter-like transiting planet (R = 1 R$_{\mbox{J}}$, P = 45 days, i = 89.5$^\circ$) with a 3 R$_{\oplus}$ moon around a star equal to the Sun. \textbf{(a)} The star has a spot with 0.5 planet radius, intensity 0.6, located at 30$^\circ$. \textbf{(b)} The star has 4 spots, with radius respectively 0.7, 0.6, 0.7 and 0.8 planet radius, intensity 0.6, 0.5, 0.6 and 0.7, located at -30, -15, 5 and 30$^\circ$. Each spot signal appears duplicated, once when the moon occults it and the other when the planet eclipses the spot.}
   \label{fig:planetas_com_lua_e_manchas}
\end{figure*}

Examples of how confusing things can get, especially how the presence of starspots may hinder the detection of moons, are shown in Figure~\ref{fig:planetas_com_lua_e_manchas}. The figure presents the transit of a planet with an orbiting moon in front of a star, in which surface there is a spot, for two situations: only one spot  (\ref{fig:planetas_com_lua_e_manchas_a}) and star with 4 spots (\ref{fig:planetas_com_lua_e_manchas_b}). In both panels, the planet is identical to Jupiter with a period of 45 d and the moon has 3 Earth radius and an orbital period of 2 d around the planet. In the case of Figure~\ref{fig:planetas_com_lua_e_manchas_b}, it is clearly seen that the presence of starspots can significantly difficult the detection of any moon effect on the light curve.

\section{Test results from the model}

To verify the detectability of moon and ring systems around a planet, it is possible to add Gaussian noise to the modeled light curve and then try and recover the input parameters from a fit. For CoRoT data simulation, we used a noise level of $6 \times 10^{-4}$, such as that observed for CoRoT-2 \citep{SilvaValio2010}, and in the case of the Kepler data, $2 \times 10^{-5}$. 

Once the noise has been added to the model light curve, we fit this light curve with the following sequence of algorithms. First the PIKAIA \citep{RefPikaia} algorithm is used to obtain a first guess that is then inputed into the AMOEBA \citep{RefAmoeba} routine. PIKAIA is a genetic algorithm the main advantage of which is to find the global maximum of the function to be maximized, using only the allowed interval value for each free parameter. These results are then used as input to the AMOEBA routine that refines the result using the simplex method {\it for convergence minimization}. The error estimates for the fitted parameters were obtained using the MPFIT algorithm \citep{MPFit}.

After the fitting procedure, we compute the {\it goodness-of-fit} using the parameter Q, as defined in \cite{NumericalRecipes} as \emph{the probability that the chi-square should exceed a particular value $\chi^2$ by chance} and calculated by:
\begin{equation}
 Q(\chi^2 | \nu) = 1 - P \left( \frac{\nu}{2}, \frac{\chi^2}{2} \right)  \;  ,
\end{equation}
where $P$ is the incomplete gamma function $P(a,x)$. Generally, if Q is larger than 0.01, then the fit is acceptable. In this case, it can be said that such event (moon or ring) is detectable. If Q is between 0.001 and 0.01, then the fit may or not be acceptable, and more studies about the noise are required. This can be caused, for example, if the measurement errors are underestimated. Q below 0.001 indicates a bad fit. On the other hand, Q very close to 1 is not expected also. This is, literally, too good to be true. This can be caused for nonnormal error distribution, or overestimated measurement errors. Next, we show some examples of the fit to light curves with noise added, and the results they yield.

\subsection{Fit of a planetary transit}

The first test to the model is to apply it to the two transiting planets described in Section~\ref{secao_transito_planeta}, where Gaussian noise has been added similar to that of CoRoT and Kepler observations. In this fit, all the stellar parameters and the orbital period (or semi-major axis) of the planet were fixed, and the only free parameters were the planetary radius and the orbital inclination angle.

\begin{figure*}[!htb]
   \centerline{
     \subfigure[ ]{\includegraphics[width=5.5cm]{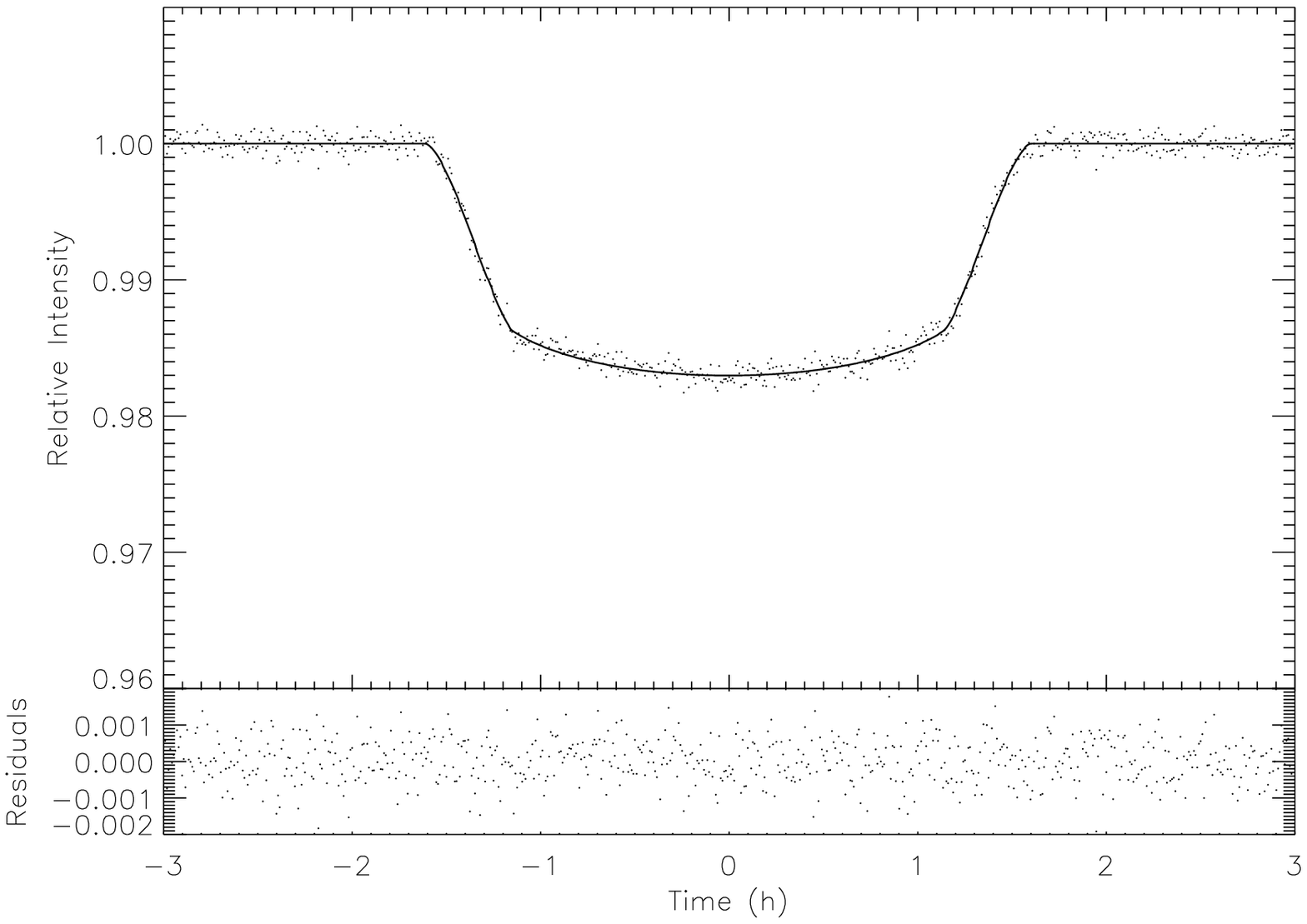}
     \label{fig:fit_planetas_sem_lua_a}}
     \hfil
     \subfigure[ ]{\includegraphics[width=5.5cm]{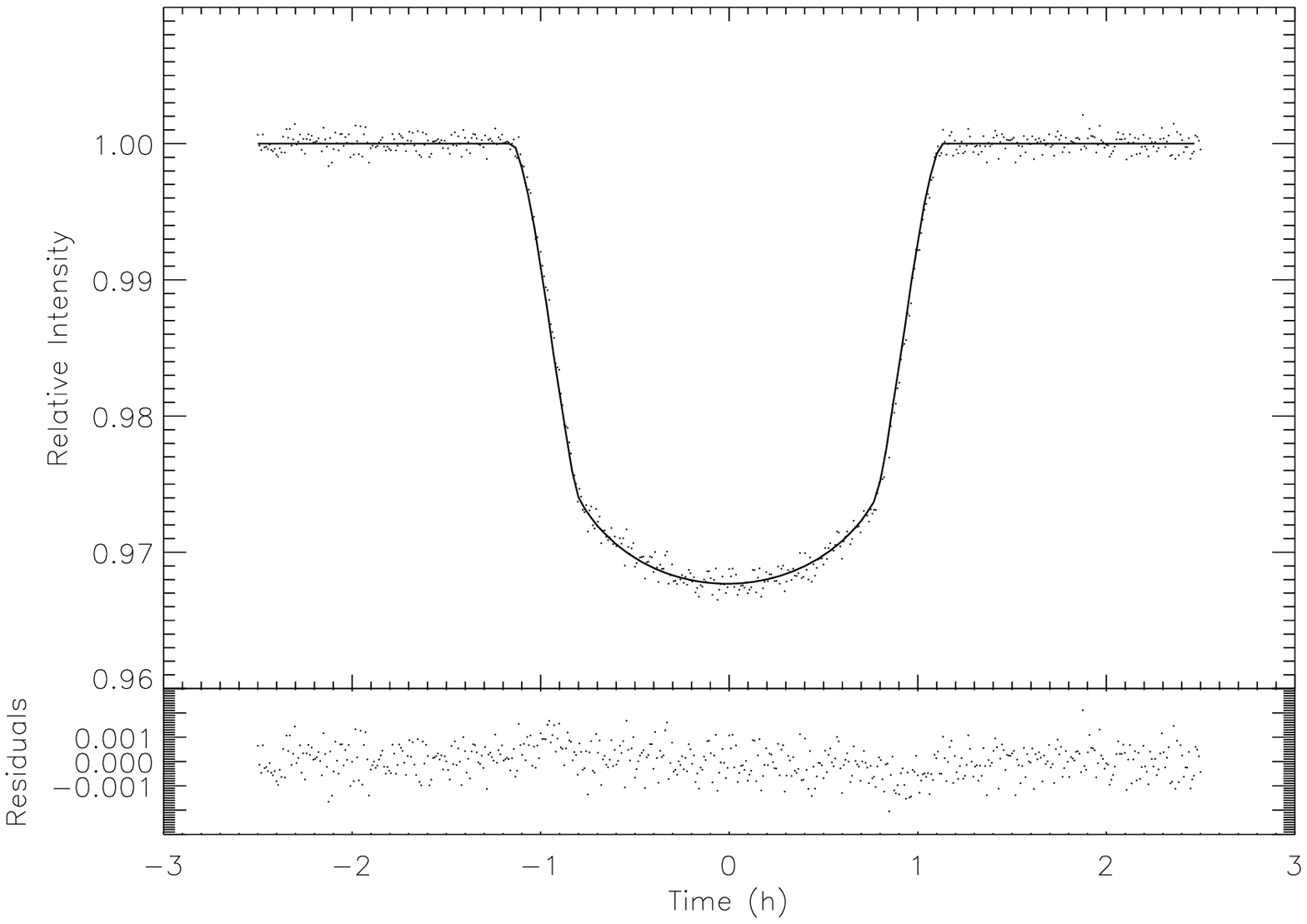}
     \label{fig:fit_planetas_sem_lua_b}}
   }
   \caption{Light curve fit of model planetary transit for \textbf{(a)} HD 209458b and \textbf{(b)} CoRoT-2b. The dots represent the noisy light curve data points, whereas the solid line shows the best yield fit.}
   \label{fig:fit_planetas_sem_lua}
\end{figure*}

The comparison between the original parameters and those obtained from the fit of the noisy modeled light curve are seen in Tables~\ref{tab:ajuste_planetas_corot} and \ref{tab:ajuste_planetas_kepler}, and and in Figures~\ref{fig:fit_planetas_sem_lua_a} and \ref{fig:fit_planetas_sem_lua_b} for HD 209458 and CoRoT-2. Both light curves had noise level similar to the CoRoT satellite. The two panels show the simulated data points as dots and the fit as a solid line. The residuals are shown in the lower part of each panel.

\begin{table}[!htb]
  \begin{center}
    \small
    \setlength{\tabcolsep}{6pt}
    \renewcommand{\arraystretch}{1.1}
    \vspace{2mm}
    \caption{Fit results for planetary transits of HD 209458b and CoRoT-2b for CoRoT simulations.}
    \begin{tabular}[c]{c|c|c|c|c|c}
      \hline\hline
      Planet & Parameter & Original Value & Fit Result & Reduced $\chi^2$ & Q \\\hline
      \multirow{2}{*}{HD 209458b} & Planet radius (R$_J$) & 1.38 & 1.382 $\pm$ 0.025    & \multirow{2}{*}{1.0639} & \multirow{2}{*}{0.098} \\
                   & Orbital inclination ($^\circ$) & 86.59  & 86.6 $\pm$ 0.3  & & \\ \cline{1-6}
      \multirow{2}{*}{CoRoT-2b}   & Planet radius (R$_J$) & 1.465 & 1.466  $\pm$ 0.016 & \multirow{2}{*}{1.0157} & \multirow{2}{*}{0.3890} \\
                   & Orbital inclination ($^\circ$) & 87.84   &  87.84 $\pm$ 0.21  & & \\ \hline
    \end{tabular}
    \label{tab:ajuste_planetas_corot}
  \end{center}
\end{table}

\begin{table}[!htb]
  \begin{center}
    \small
    \setlength{\tabcolsep}{6pt}
    \renewcommand{\arraystretch}{1.1}
    \vspace{2mm}
    \caption{Fit results for planetary transits of HD 209458b and CoRoT-2b for Kepler simulations.}
    \begin{tabular}[c]{c|c|c|c|c|c}
      \hline\hline
      Planet & Parameter & Original Value & Fit Result & Reduced $\chi^2$ & Q \\\hline
      \multirow{2}{*}{HD 209458b} & Planet radius (R$_J$) & 1.38 & 1.3797  $\pm$ 0.0011 & \multirow{2}{*}{1.0627} & \multirow{2}{*}{0.1747} \\
                   & Orbital inclination ($^\circ$) & 86.59  & 86.594 $\pm$ 0.015 & & \\ \cline{1-6}
      \multirow{2}{*}{CoRoT-2b}   & Planet radius (R$_J$) & 1.465& 1.4636 $\pm$ 0.0007 & \multirow{2}{*}{1.1156} & \multirow{2}{*}{0.0859} \\
                   & Orbital inclination ($^\circ$) & 87.84   &  87.839  $\pm$ 0.009& & \\ \hline
    \end{tabular}
    \label{tab:ajuste_planetas_kepler}
  \end{center}
\end{table}

\subsection{Transit fit of a planet with moon}

If the planet has a moon in orbit around it, there are two factors that make its detection difficult. The first one is the noise, as expected. In cases where the moon is very small, their signature in the light curve is flooded by the noise level, leaving it undetectable. Another difficulty is the presence of starspots on the stellar surface, mentioned above.

\begin{figure}[!htb]
 \centering
 \includegraphics[width=7.5 cm]{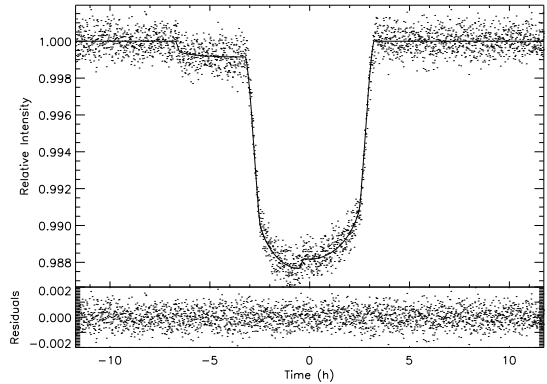}
 \caption{Light curve fit for the first transit of a hot Jupiter (R = 1 R$_{\mbox{J}}$, P = 45 days, i = 89.5$^\circ$) with a moon (R$_{\mbox{M}}$ = 3 R$_{\oplus}$, P$_{\mbox{M}}$ = 5 days) considering the noise level of CoRoT. }
 \label{fig:fit_planeta_lua_corot}
\end{figure}

To demonstrate how moons can be detected, we consider the case of a hot Jupiter with a period of 45 days around its host star, modelled with no spots. In orbit around this planet there is a moon with 3 Earth radii and a period of 5 days. Simulations were performed for both noise levels of CoRoT and Kepler. The parameters obtained from the fit and the original ones are listed in Table~\ref{tab:ajuste_planeta_lua}, whereas the fit results are plotted in Figure~\ref{fig:fit_planeta_lua_corot}, for CoRoT noise level. The fit was performed for 5 consecutive transits, but the figure shows only the first.

\begin{table}[!htb]
  \begin{center}
    \small
    \setlength{\tabcolsep}{6pt}
    \renewcommand{\arraystretch}{1.1}
    \vspace{2mm}
    \caption{Fit result of 5 transits of a planet with an orbiting moon.}
    \begin{tabular}[c]{c|c|c|c|c|c}
      \hline\hline
      Noise Level             & Parameter               & Original Value & Fit Result & Reduced $\chi^2$ & Q \\ \hline
      \multirow{4}{*}{CoRoT}  & Planet radius (R$_J$) & 1.0     &   0.9976 $\pm$ 0.0005 & \multirow{4}{*}{1.0070} & \multirow{4}{*}{0.3954} \\
                  & Moon period (days)                & 5.0     &   5.001 $\pm$ 0.012 & & \\
                  & Moon radius (R$_\oplus$)             & 3.0     &   2.979 $\pm$ 0.019& & \\
                  & Moon position angle ($^\circ$)   & 0.0     &   0.21 $\pm$ 0.11& & \\ \hline
      \multirow{4}{*}{Kepler} & Planet radius (R$_J$) & 1.0     &   0.9999 $\pm$ 0.0007& \multirow{4}{*}{1.0419} & \multirow{4}{*}{0.1337} \\
                  & Moon period (days)                & 5.0     &   5.000 $\pm$ 0.023& & \\
                  & Moon radius (R$_\oplus$)             & 3.0     &   2.999 $\pm$ 0.024& & \\
                  & Moon position angle ($^\circ$)   & 0.0     &   -0.4 $\pm$ 0.6& & \\ \hline
    \end{tabular}
    \label{tab:ajuste_planeta_lua}
  \end{center}
\end{table}

We noticed that the larger the number of transits considered in the fit, the smaller the number of iterations necessary for the fitting algorithms to obtain a good fit, that is, the easier it is to detect the moon.

\subsection{Transit fit of a planet with rings}

As seen above, planetary rings can be detected by the round shape of the transit at ingress and egress times, besides the larger depth of the transit (due to the increase in the stellar surface area covered). For these signatures to be detectable, the noise is the predominant factor. If the signature is too small, the noise will completely overcome it.

\begin{figure*}[!htb]
   \centerline{
     \subfigure[ ]{\includegraphics[width=7.5cm]{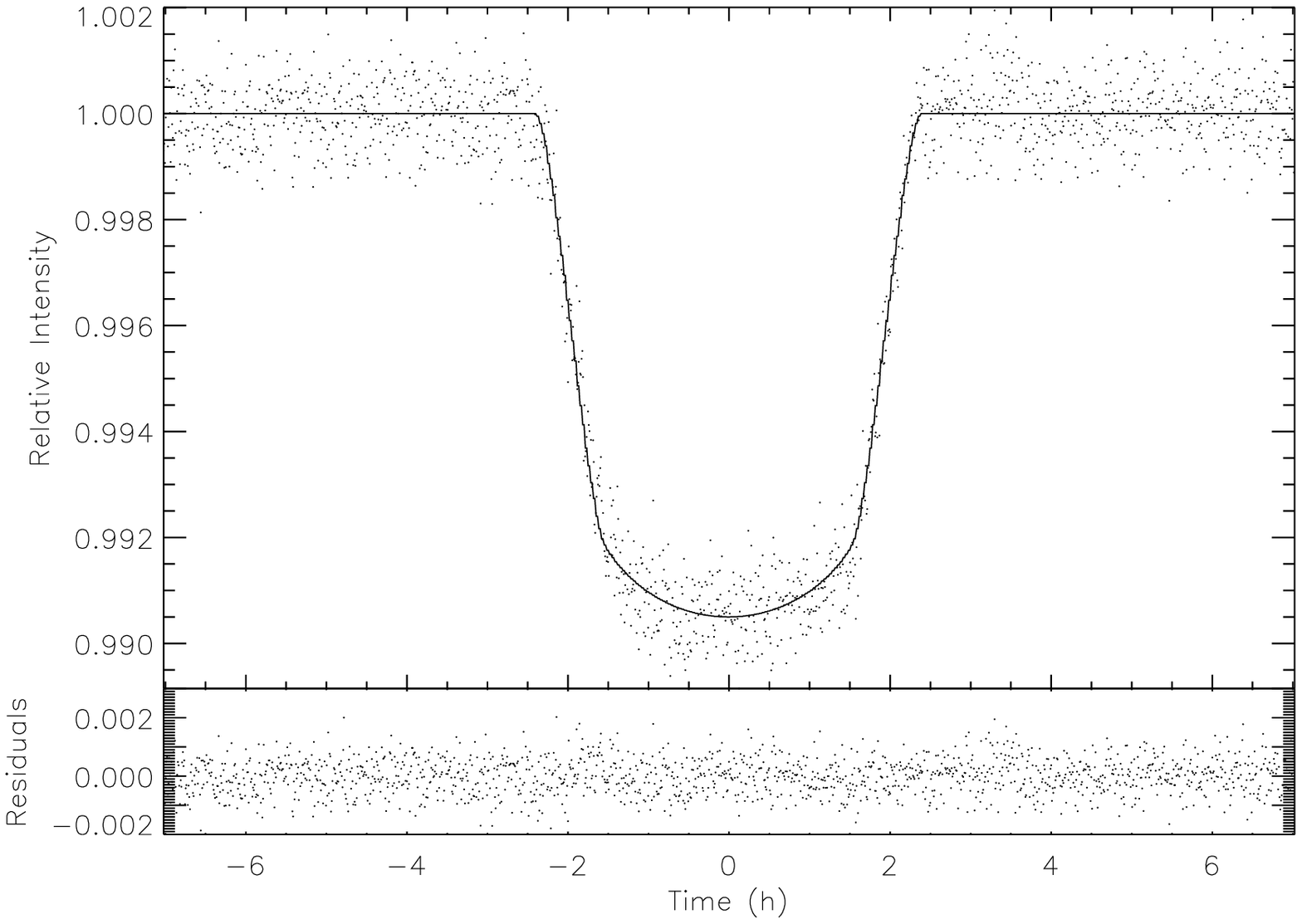}
     \label{fig:saturno_planeta_corot}}
     \hfil
     \subfigure[ ]{\includegraphics[width=7.5cm]{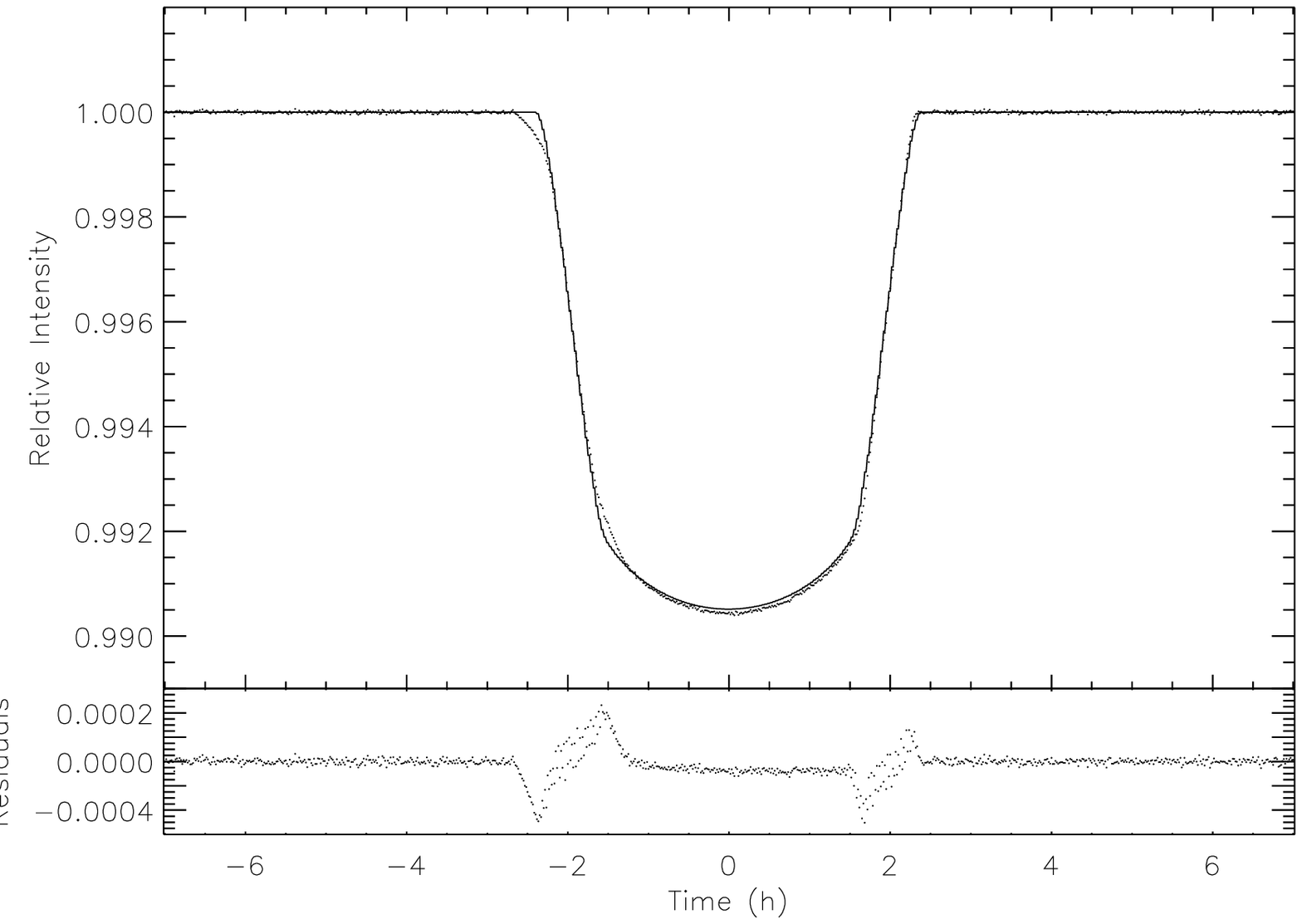}
     \label{fig:saturno_planeta_kepler}}
   }
   \caption{Saturn-like planet (R = 0.84 R$_{\mbox{J}}$, P = 5 days, i = 88$^\circ$) transit fitted without considering the presence of its rings for noise levels of the \textbf{(a)} CoRoT and\textbf{(b)} Kepler telescopes, respectively.}
   \label{fig:saturno_planetas}
\end{figure*}

To exemplify, we simulate the transit of a planet identical to Saturn and its rings, with 30-days period and 89$^\circ$ orbital inclination. The transit is modeled for two cases: that with the noise level of CoRoT and the other with that of Kepler telescope. In both cases, the light curve was submitted to a fit process assuming only the presence of the planet. The result of this preliminary fit is shown in Figure~\ref{fig:saturno_planetas} and Table~\ref{tab:ajuste_aneis_planetas}.

\begin{table}[!htb]
  \begin{center}
    \small
    \setlength{\tabcolsep}{6pt}
    \renewcommand{\arraystretch}{1.1}
    \vspace{2mm}
    \caption{Fit results for a Saturn-like planet simulation, considering only the planet.}
    \begin{tabular}[c]{c|c|c|c|c|c}
      \hline\hline
      Noise level & Parameter& Original Value & Fit result & Reduced $\chi^2$ & Q \\\hline
      \multirow{2}{*}{CoRoT} & Planet radius (R$_J$) & 0.84 & 0.934 $\pm$ 0.022 & \multirow{2}{*}{1.0291} & \multirow{2}{*}{0.2050} \\
                             & Orbital inclination ($^\circ$) &  89  & 89.0 $\pm$ 0.4& & \\ \hline
      \multirow{2}{*}{Kepler}& Planet radius (R$_J$) & 0.84  & 0.9434 $\pm$ 0.0009& \multirow{2}{*}{25.10} & \multirow{2}{*}{0} \\
                             & Orbital inclination ($^\circ$) & 89  & 88.999 $\pm$ 0.019   & & \\ \hline
    \end{tabular}
    \label{tab:ajuste_aneis_planetas}
  \end{center}
\end{table}

From the analysis of the residual plotted on Figure~\ref{fig:saturno_planetas} and the Q value in Table~\ref{tab:ajuste_aneis_planetas}, we see that the fit considering only the planet yields a satisfactory result only for the CoRoT noise level (Figure~\ref{fig:saturno_planeta_kepler}). This indicates that a ring system such as that of Saturn could only be detected by the Kepler satellite. The CoRoT satellite would identify this light curve as that produced by a giant planet with a radius slightly larger than that of Saturn (12.3 \%), but without rings.

\begin{figure}[!htb]
 \centering
 \includegraphics[width=7.5 cm]{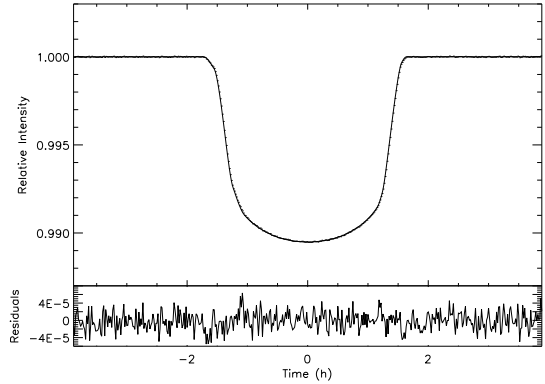}
 \caption{Saturn-like planet (R = 0.84 R$_{\mbox{J}}$, P = 5 days, i = 88$^\circ$) transit fitted, considering now the presence of rings.}
 \label{fig:fit_saturno_kepler}
\end{figure}

Next, we fit only the light curve plotted in Figure~\ref{fig:saturno_planeta_kepler} but now considering the presence of rings, that is the ring parameters are taken into account. The resulting fit parameters are listed in Table~\ref{tab:ajuste_saturno_kepler}, whereas the result of the fit is plotted  in Figure~\ref{fig:fit_saturno_kepler}. Comparing the residual of Figures~\ref{fig:saturno_planeta_kepler} and \ref{fig:fit_saturno_kepler}, one can see that the fit is much better for the model where the rings were included.

\begin{table}[!htb]
  \begin{center}
    \small
    \setlength{\tabcolsep}{6pt}
    \renewcommand{\arraystretch}{1.1}
    \vspace{2mm}
    \caption{Resulting parameters for the case of a transiting Saturn, as would be observed by Kepler.}
    \begin{tabular}[c]{c|c|c|c|c}
      \hline\hline
      Parameter                     & Original Value & Fit Result & Reduced $\chi^2$ & Q\\\hline
      Planet radius (R$_J$)               & 0.84   & 0.839 $\pm$ 0.005   & \multirow{6}{*}{1.0173} & \multirow{6}{*}{0.3553} \\
      Orbital inclination ($^\circ$)      & 89     & 89.00 $\pm$ 0.03 & & \\
      Inner radius (R$_P$)                  & 1.11   & 1.11 $\pm$ 0.06   & &\\
      Outer radius (R$_P$)                  & 2.32   & 2.316 $\pm$ 0.026   & &\\
      Angle with the sky plane ($^\circ$)    & 78.0   & 77.9 $\pm$ 1.5  & &\\
      Angle with the orbital plane ($^\circ$)   & 20.0   & 19.9 $\pm$ 0.5 & &\\
      Transparency                         & 0.5    & 0.51 $\pm$ 0.06   & &\\ \hline

    \end{tabular}
    \label{tab:ajuste_saturno_kepler}
  \end{center}
\end{table}

The Q value in Table~\ref{tab:ajuste_saturno_kepler} and the residuals in Figure~\ref{fig:fit_saturno_kepler}  show that the fit was satisfactory.

\section{Detection limits}

To determine the radius of the smallest detectable moon, several simulations were performed decreasing the radius at each run. Every simulation had its light curve fitted yielding the moon parameters. Whenever the Q value indicates a good fit, we consider the moon detectable. Thus, in the case of CoRoT, the smallest detectable moon in orbit around a Jupiter size planet has a radius of 1.3 R$_\oplus$, while in the case of Kepler, this minimum radius decreases to 0.3 R$_\oplus$. The latter moon would be smaller than Ganimedes, the larger moon of our Solar System, which size is 0.41  R$_\oplus$. In all the simulations, the star was considered Sun-like with no spots. 
 
For the case of planetary rings, the process is slightly different, because, in this case, the biggest problem is distinguishing between single planets and those with rings. In this case, then, we adopt a criterion that takes into consideration the residuals. When fitting the light curve of a planet with rings, first only the planet parameters are fit. If the Q value indicates a bad fit, then we may conclude that the rings can be detectable. 
 
To establish a detectability criterion, first we define the effective area of the rings as the ring projection on the plane of the sky multiplied by its opacity, according to:

\begin{equation}
A_{eff} = \left( R_E^2 - R_I^2 \right) \pi \cos{\psi} \left( 1 - \tau \right)
\end{equation}

\noindent where $R_I$ and $R_E$ are the internal and external radii, respectively,  $\psi$ is the inclination angle of the rings relative to the plane of the sky, and $\tau$ is the transparency. Totally transparent rings have $\tau=1$, whereas totally opaque ones have a zero value of transparency.

The results of our simulations showed that in the case of the Kepler data, planetary rings are detectable when the effective area corresponds to approximately 3\% of the area covered by the planet. For example, rings with $\tau=0.5$ and inclination angle of 78$^\circ$ with the plane of the sky in orbit a Saturn-like planet, will be detectable with rings about 51\% the size of those of Saturn. For the CoRoT data, the effective area corresponds to approximately the area of the planet. Therefore, for the same planet as in the previous case, the rings should have sizes 45\% larger than those of Saturn to be detectable above the CoRoT noise level.

\section{Conclusion}

Here we reported the results of a model capable of simulating planetary transits, including the effect of existing moons and rings around that planet. Light curves were presented for each case. Estimates of the threshold detection for such events were made. It was shown how these events may be detected using the PIKAIA and AMOEBA routines for curve fitting. As expected, the main limiting factor for the possible detection is the noise level of the telescopes. This can be seen by comparing the results from simulations of photometric light curves of CoRoT and Kepler.

The CoRoT data allows for the detection of moons with at least 1.3 R$_\oplus$ in size around a planet identical to Jupiter. In the case of Kepler data, this detection limit on the moon radius drops to  0.3 R$_\oplus$. For planetary rings, a Saturn-like planet would be detectable with Kepler, however, this same planet-ring system would be mistakenly inferred as a single planet with slightly larger radius by CoRoT. Detection limits for both telescopes were estimated above.

Next, this model will be applied to real light curves to search for the types of signatures described in this work. Data from both CoRoT and Kepler will be used. An improvement that can be made to the model is the implementation of analysis of the transit time and duration variations caused by the presence of the moon that makes the planet orbit the common center--of--mass \citep{Kipping2009, Kipping2009b}. These timing variations would confirm the presence of moons. Finally, by detecting the moon by both methods, photometry and timing variation, it is possible to establish a complete set of physical parameters for the satellite.

\section{Acknowledgments}

The authors acknowledge financial support from the Brazilian agencies CAPES and FAPESP (grant number 2006/50654-3).

\end{document}